\documentclass[fleqn,10pt]{wlscirep}
\usepackage[utf8]{inputenc}
\usepackage[T1]{fontenc}
\usepackage{hyperref}
\title{Tracking Blobs in the Turbulent Edge Plasma of a Tokamak Fusion Device}

\author[1,*]{Woonghee Han}
\author[2]{Randall A. Pietersen}
\author[2]{Rafael Villamor-Lora}
\author[3]{Matthew Beveridge}
\author[4]{Nicola Offeddu}
\author[1]{Theodore Golfinopoulos}
\author[4]{Christian Theiler}
\author[1]{James L. Terry}
\author[1]{Earl S. Marmar}
\author[3]{Iddo Drori}

\affil[1]{MIT Plasma Science and Fusion Center, Cambridge, Massachusetts 02139, USA}
\affil[2]{MIT, Civil and Environmental Engineering, Cambridge, Massachusetts 02139, USA}
\affil[3]{MIT, Electrical Engineering and Computer Science, Cambridge, Massachusetts 02142, USA}
\affil[4]{École Polytechnique Fédérale de Lausanne (EPFL), Swiss Plasma Center (SPC), CH-1015 Lausanne, Switzerland}

\affil[*]{harryhan@mit.edu}

\begin{abstract}
The analysis of turbulence in plasmas is fundamental in fusion research. Despite extensive progress in theoretical modeling in the past 15 years, we still lack a complete and consistent understanding of turbulence in magnetic confinement devices, such as tokamaks. Experimental studies are challenging due to the diverse processes that drive the high-speed dynamics of turbulent phenomena. This work presents a novel application of motion tracking to identify and track turbulent filaments in fusion plasmas, called blobs, in a high-frequency video obtained from Gas Puff Imaging diagnostics. We compare four baseline methods (RAFT, Mask R-CNN, GMA, and Flow Walk) trained on synthetic data and then test on synthetic and real-world data obtained from plasmas in the Tokamak à Configuration Variable (TCV). The blob regime identified from an analysis of blob trajectories agrees with state-of-the-art conditional averaging methods for each of the baseline methods employed, giving confidence in the accuracy of these techniques. High entry barriers traditionally limit tokamak plasma research to a small community of researchers in the field. By making a dataset and benchmark publicly available, we hope to open the field to a broad community in science and engineering.
\end{abstract}

\begin{document}

\flushbottom
\maketitle

\thispagestyle{empty}

\section*{Introduction}
Due to the enormous quantity of energy released by the fusion reaction, the virtually inexhaustible fuel supply on earth, and its carbon-free nature, nuclear fusion is a highly desirable energy source with the potential to help reduce the adverse effects of climate change. Fusion research is, therefore, an ongoing worldwide effort \cite{becoulet2021,ball2021}.

\begin{figure}[ht]
 \phantomsection
 \centering
 \includegraphics[width=0.95\textwidth]{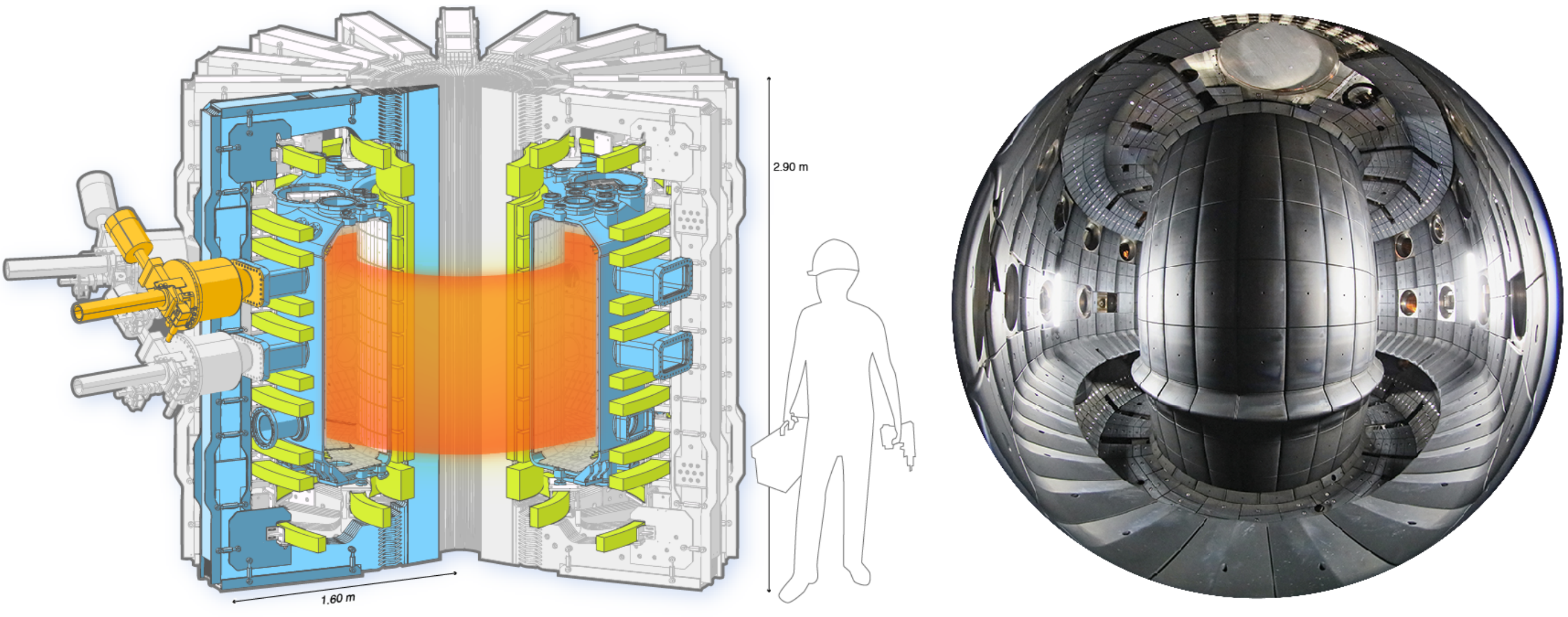}
 \caption{The schematic of the Tokamak à Configuration Variable (TCV) (left) and its interior (right). Credits: EPFL (left) and A. Herzog, EPFL (right).}
 \label{fig:tcv_fig}
\end{figure}

In order to approach the conditions necessary for sufficient fusion reactivity, the fuel -- a mixture of heavy isotopes of hydrogen -- must be raised to extremely high temperatures, above 100 million degrees Celsius \cite{tester2012,becoulet2021} -- for comparison, the core of the sun is roughly 15 million degrees Celsius. Under these conditions, the fuel, like all stars, is in the plasma state and must be isolated from material surfaces. Several confinement schemes have been explored over the past 70 years \citation{stacey2010}.  Of these, the tokamak device, a scheme first developed in the 1950s, is the best-performing fusion reactor design concept to date \citation{wurzel2022}.  It uses powerful magnetic fields of several to over 10 Tesla to confine the hot plasma -- for comparison, this is several times the field strength of magnetic resonance imaging machines (MRIs). The Tokamak à Configuration Variable (TCV) \cite{Reimerdes_2022}, sited in Lausanne, Switzerland and shown in Figure \ref{fig:tcv_fig}, is an example of such a device and provides the data presented here.

The research addressed in this paper involves phenomena that occur around the boundary of the magnetically confined plasma within TCV. The boundary is where the magnetic field-line geometry transitions from being ``closed" to ``open ."The ``closed" region is where the field lines do not intersect material surfaces, forming closed flux surfaces. The ``open" region is where the field lines ultimately intersect material surfaces, resulting in a rapid loss of the particles and energy that reach those field lines. Around this boundary (called the Last Closed Flux Surface or LCFS) is a region of enhanced turbulent transport across the field lines (``perpendicular" transport). Therefore, the transport occurring here is of great interest because of its role in plasma confinement and dealing with the exhaust heat and particle loads on the intersected material surfaces. 

\begin{figure}[ht]
  \phantomsection
  \centering
  \includegraphics[width=1.0\textwidth]{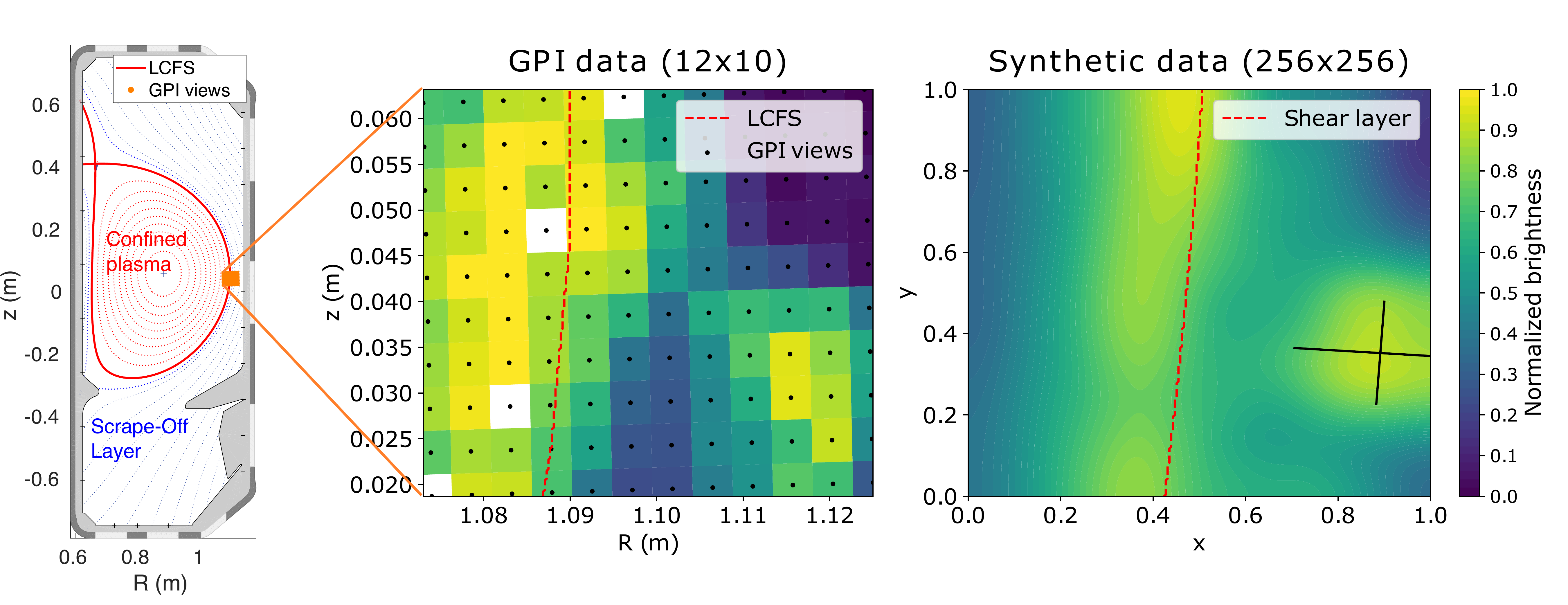}
  \caption{Cross-section of a plasma in the Tokamak à Configuration Variable (TCV) with the locations of Gas Puff Imaging (GPI) views near the Last Closed Flux Surface (LCFS) (left). Snapshot of experimental GPI data capturing a blob on the right-hand side moving radially outward (middle). Here empty (white) spots correspond to dead GPI views. The brightness level is coded in the color bar to the right, with low as blue and high as yellow. Snapshot of synthetic data capturing a blob moving radially outward (right). A Gaussian ellipse represents the blob with a major and minor axis marked by perpendicular black lines.}
  \label{fig:tcv_GPI_syntehtic}
\end{figure}

Researchers use a technique called Gas-Puff Imaging (GPI) \cite{Zweben_2017} to visualize the phenomena occurring at and around the plasma boundary in both space and time. A small amount of neutral gas is locally injected into the region of interest. The visible-light emissions resulting from the interaction of the plasma with this gas cloud are captured along sight lines that are tangential to the local magnetic field. \textit{Analysis of the time sequences of the images produced by this technique is the primary subject of this work.} At TCV, the 2D GPI data are collected at frame rates of $\sim$ 2 MHz, significantly faster than the turbulence timescales of interest. The images represent what is occurring perpendicular to the direction of the local magnetic field lines. This is illustrated schematically in Figure \ref{fig:tcv_GPI_syntehtic} (left), showing cross-sections of plasma flux surfaces and the TCV vacuum vessel. (The center-line of this toroidally symmetric plasma geometry is to the left of the figure.) A $\sim$50$\times$40~mm cross-section of plasma, spanning the LCFS at the outer edge of the plasma, is imaged with a 12$\times$10 pixel array. The images' motion from left to right increases the radial coordinate and moves away from the closed flux surfaces. The most prominent features in the image sequences are 1) a bright band of emission that is aligned with the LCFS and 2) the expulsion of features from the LCFS region that are brighter than their surroundings and the subsequent motion of these features. In the literature, these features are commonly called ``filaments'' (because they are actually extended in the third dimension, along the magnetic field lines) or ``blobs'' (because of their appearance in cross-section in the images) \cite{DIppolito_2011}. These filaments are strong perturbations (of order 1) in the local plasma conditions, with elevated electron temperature and density \cite{Agostini_2015,Kube_2018}. They have typical auto-correlation times of roughly 10 $\mu$s and, therefore, can exist as distinct features for multiple frames, but their shape and intensity can gradually change frame-to-frame. \textit{In this work, we are mainly interested in the detection and tracking of these ``blob'' features.}

The primary problem we are addressing in this paper is how to process these image sequences in ways that will ultimately allow estimation of the contributions of this blob transport to the heat and particle fluxes leaving the confined plasma. We are also addressing how to process the images to facilitate continued and more detailed comparisons of the experimental observations with theoretical models of blob dynamics. The theoretical models predict scalings for the relation between blob size and radial velocity depending on key plasma parameters, such as local temperature and collisionality \cite{Myra_2006_1,Myra_2006_2}.

This primary problem is essential since the expulsion and transport of these objects represent particle and energy loss from the confined plasma. This has enormous consequences for the interaction of the hot plasma with the plasma-facing material surfaces that are in place to accept the exhaust heat and particles. Parallel heat fluxes of hundreds of MW/m$^2$ are obtained in the narrow exhaust channels of present-day devices \cite{Brunner_2018}, and the parallel heat fluxes will be even higher in the near-future net-energy-gain tokamaks like SPARC \cite{Kuang_2020} and ITER \cite{Goldston_2015}. The radially-outward motion of the blobs can broaden this exhaust channel, thereby reducing the peak heat- and particle fluxes, the desired circumstance. However, the same motion can also increase an undesired plasma interaction with the other plasma-facing components. \textit{These considerations constitute the highest-level motivations for an accurate assessment of the blobs' occurrence, size, and motion.} Combining knowledge of the occurrence frequency, size, and radial velocity with temperature and density within the blobs allows quantitative estimations of radial particle- and energy fluxes. The additional problem that we are addressing, facilitating closer experimental comparisons with theoretical models of blob transport, is essential for experimental validation tests of these models. These tests of the models (e.g., \cite{Offeddu_2022}) are made more valuable with accurate determinations of the size, shape, and motion of blobs.

Traditional approaches to blob analysis include a family of conditional averaging methods, custom-made workflows that track high signal regions, and spatio-temporal cross-correlation techniques \cite{Offeddu_2022,Zweben_2017,tracking}. The conditional averaging methods and the cross-correlation techniques have the limitations of only providing averaged characteristics of blobs. The custom-made workflows are non-standardized and are not benchmarked. The data sets that we are working with typically include $\sim10^5$ sequential images, having been recorded over a time-subset shorter than the $\sim1$~s plasma pulse on TCV. 30 such plasma discharges can be produced per each working day at TCV; hence this data throughput is too high for human by-eye analysis. Machine learning allows accurate and efficient analysis of turbulence characteristics utilizing these extensive datasets.

This work presents a novel application of four well-known, standardized, and benchmarked tracking methods to track blobs in GPI images. These methods are trained to reproduce identification of blobs by humans as close as possible since human subjectivity isolates blobs from the non-blobs in the GPI images. We evaluate their performance by comparing them to the brightness-threshold contours and human detection/tracking using a limited set of real-data images. In \hyperref[sec:methods]{Methods} we describe the four models: 1) optical flow detection using Recurrent All-pairs Field Transforms (RAFT) \cite{teed2020raft}; 2) mask detection using Mask Region-based Convolutional Neural Network (Mask R-CNN) \cite{He_2018} in combination with Bayesian optimization \cite{balandat2020botorch} for tuning the hyperparameters in the training; 3) optical flow detection using Global Motion Aggregation (GMA) \cite{jiang2021learning}; and 4) optical flow detection using Flow Walk \cite{bian2022flowwalk,bian2022learning}. In \hyperref[sec:methods]{Methods} we also define the performance metrics and the workflow. The performance comparisons are given in \hyperref[sec:results]{Results}, as is a description of the results obtained by human detection, which shows some ambiguity in what constitutes a blob in real experimental data. Finally, the trained models are applied to an active plasma-physics research topic, i.e., finding the regime of the blob dynamics, as described in Myra \emph{et al.} \cite{Myra_2006_1,Myra_2006_2}, and the models' results are compared with the result from a Conditional Average Sampling (CAS) method that has been used previously for this purpose \cite{Offeddu_2022}. The work is summarized in \hyperref[sec:conclusion]{Conclusions}.  

\phantomsection
\section*{Methods}
\label{sec:methods}

We train our models on synthetic data and test them on synthetic and real data. Synthetic data affords several advantages for training a model. First, our models use optical flow vectors (RAFT, GMA, Flow Walk) or masks (Mask R-CNN) of the blob objects as ground-truth features that humans cannot adequately label. The user has perfect knowledge of and control over these features and noise in the synthetic dataset. By contrast, feature recognition in available real data may be prohibitively costly to process, hard to reproduce, poor quality, and not representative of general phenomenology. Second, synthetic datasets of arbitrary size and complexity can be made more generic than a small, real, vetted dataset, such that a model may be trained to recognize features that do not appear in available real data but might appear more broadly. There are a few edge cases in the real data to be covered in training. Events such as splitting or merging blobs often occur but not often in real data. We further iterate upon the generation of synthetic data by adding, removing, or emphasizing characteristics like these events. Moreover, synthetic data is more efficient in acquisition and storage than real data. Real GPI data is costly to acquire; blob dynamics vary with the plasma condition, and we need to run many experiments to generate a significant variety of blob dynamics in the real world. Also, the algorithm to create the synthetic data is more compact than a large real dataset and is easily adapted to other situations where the effort expended in processing the real dataset cannot be reused. Therefore, by using synthetic data, we can have inexpensive, variable, and unlimited data generation \cite{nikolenko2021synthetic,syntheticdata4cv2022cvprtutorial,syntheticdata4ml2022cvprworkshop}.

An essential goal of this contribution is to solicit greater participation in fusion research from the broader machine-learning community. Toward this end, we make our synthetic training dataset and a real dataset available to benchmark performance against other models. We hope this may inspire readers not only to evaluate their model performance against this fusion-relevant task but to seek out and engage more generally to help solve critical problems in the field of fusion energy.

\subsection*{Dataset}

\paragraph{Real Data}
Each experimental GPI video is a length $t$ series of grayscale images with $12\times10$ pixels, where $t$ is the number of frames in the video. A snapshot of an experimental GPI video is shown in Figure \ref{fig:tcv_GPI_syntehtic} (middle), which shows a blob (as a bright spot on the bottom-right), and the Last Closed Flux Surface (LCFS), which is also approximately the position of a shear layer, across which the vertical plasma background flow reverses direction. Before inputting the experimental GPI data into the models, we standardize the amplitude ranges between 0 and 1 and remaps the images to a finer spatial grid. We subtract the mean brightness from the individual pixel brightnesses and divide the result by the standard deviation. It is also clipped by $\frac{1}{n}\sum_{t=1}^{n} C_{min}\left(t\right)$ and $\frac{1}{n}\sum_{t=1}^{n} C_{max}\left(t\right)$, where $n$ is the total number of frames, and $C_{min}\left(t\right)$ and $C_{max}\left(t\right)$ are the minimum and maximum (standardized) brightnesses of the frame at time $t$, respectively. In addition, although the GPI sensors output data with $12\times10$ pixels per frame, the images become better interpreted by the models when we upsample their resolution, in the present case, to $256\times256$ pixels per frame. We use radial basis function (RBF) interpolation with a cubic function. This degree of upsampling allows adequate detail during analysis without being too computationally demanding. We design the models to receive inputs and produce outputs at this resolution.

\paragraph{Synthetic Data}
We maintain the same upsampling convention as real data by generating synthetic data for training at $256\times256$ pixels per frame. We approximate blobs in the synthetic data as ellipses, and a blob's size and speed at each moment are randomized so that a random process forms the trajectory. Here, the ellipse boundary is chosen to be the full width at half maximum (FWHM) of the brightness since the blob sizes are conventionally estimated as the FWHM of the density perturbation \cite{Fuchert_2016} which is approximately proportional to the brightness in the GPI data. Besides blobs, we simulate background flows as slowly moving, elongated ellipses. Figure \ref{fig:tcv_GPI_syntehtic} (right) shows a snapshot of such a synthetic video. We also simulate complex instances in the real GPI data for these synthetic data, such as merging and splitting blobs. The brightness of the synthetic data ranges between 0 and 1. We save the mask of blobs and the velocities at pixels in blobs as an optical flow image for every frame. We use these images as labels for training.

\subsection*{Benchmark}
Our solution combines optical flow/mask detection with \hyperref[subsubsec:tbdalgo]{Tracking-by-detection} to achieve accurate identification and tracking of the blobs. We provide and compare four baselines for detecting blobs and measuring performance using standard and specialized metrics.

\subsubsection*{Baselines}

\paragraph{RAFT} Recurrent all-pairs field transforms (RAFT) \cite{teed2020raft} computes optical flow by extracting features of pixels and building multiscale 4D correlation volumes for all pairs of pixels. RAFT iteratively updates an optical flow field using a recurrent unit that uses the correlation volumes.

\paragraph{Mask R-CNN with BO} We applied mask region-based convolutional neural network (mask R-CNN) \cite{He_2018}, which computes segmented masks of the data. Bayesian optimization is used to find the hyperparameter values that minimize the loss of the mask R-CNN during the training. We apply Bayesian optimization \cite{balandat2020botorch} with two levels of hyperparameters: (1) learning rate, momentum, weight decay, and number of epochs; and (2) data augmentation transformations $P_{horizontalFlip}$, $P_{scale}$, $P_{translate}$, $P_{shear}$, $P_{rotate}$, as well as dropout probability $P_{dropout}$ of each dropout layer in the model. The exploration ranges for each hyperparameter are in the supplementary information. We first optimize the group one hyperparameters and then keep those values fixed while optimizing the second group of hyperparameters.

\paragraph{GMA} We also implemented global motion aggregation (GMA) \cite{jiang2021learning} to estimate hidden motions. It finds long-range dependencies between pixels in an image and performs global aggregation of corresponding motion features using a transformer model. GMA is tailored to perform well on occluded regions which is not the case in our application, so this feature is not exploited. Nonetheless, we implemented GMA for comparison.

\paragraph{Flow Walk} Flow Walk learns pixel trajectories with multiscale contrastive random walks by computing the transition matrix between frames in a coarse-to-fine manner \cite{bian2022flowwalk,bian2022learning}. Flow Walk works well on detecting pixel-level changes of objects with high spatial frequencies, which is not the feature that appears on our data, as we are tracking smooth blobs. As with GMA, we implemented Flow Walk for comparison.

\phantomsection
\subsubsection*{Tracking-by-detection}
\label{subsubsec:tbdalgo}

After detecting blobs in each frame, temporal coherence between frames is enforced based on a tracking-by-detection workflow (illustrated in the supplementary information) using any of the baselines, which consists of four steps:

\begin{enumerate}

  \item \textbf{\textit{Object detection from the model}}. Given the input image sequence, the masks of blobs are predicted for each frame by the model. 
  
  \item \textbf{\textit{Feature extraction}}. We select the objects to be tracked (i.e., blob objects) by discarding predictions with scores below a threshold.
  
  \item \textbf{\textit{Pairwise cost}}. We exploit the temporal coherence of the images by computing the pairwise cost between the objects in the current and previous frames using the VIoU cost metric.
  
  \item \textbf{\textit{Bipartite matching}}. Using the cost matrix from the previous step, we assign unique correspondence between objects with the constraint that no object receives more than one assignment. If a new object appears in an isolated frame (i.e., it has no correspondence either in the previous or the next frame), it is ignored and discarded as a noise fluctuation. When a new object appears in the current frame with no correspondence in the previous frame but with a match in the next frame, we start a new track. By keeping track of the active and finished tracks, we assign IDs to the blobs in the video and record their trajectories.
\end{enumerate}

\phantomsection
\subsubsection*{Performance metrics}
\label{subsubsec:metrics}
We use standard and specialized metrics to evaluate performance.

\begin{itemize}
\item Endpoint error (EPE): The standard error measure for evaluating optical flow is defined by $\frac{1}{n}\sum \|\vec{v}_{est}-\vec{v}_{gt}\|$, where $n$ is the number of pixels in the image, $\vec{v}_{est}$ is the estimated flow field and $\vec{v}_{gt}$ is the proxy ground truth flow field.

\item Volumetric IoU (VIoU): Object detection algorithms usually use intersection over union (IoU) metrics to evaluate the prediction quality. For the prediction/ground-truth pair (e.g., bounding boxes, mask), the IoU is computed as the ratio between the intersection and union areas. While this approach is quite robust when dealing with the mask detection of solid objects with well-defined, sharp boundaries, IoU can mislead our application's score when the intersection area has a low brightness level, which is intuitively a poor prediction. Note that the boundary definition of blob features within the images is a function of the brightness threshold. Since this threshold may vary from experiment to experiment, in our case, it is more appropriate to describe the blob shape as a volume, with volumetric IoU (VIoU), using the brightness level for the third axis (illustrated in the supplementary information), and defined as:
\begin{equation}
\phantomsection
\label{eq:viou}
VIoU=\frac{\sum_{x,y\in A_{intersection}} B_{x, y}}{\sum_{x,y\in A_{union}} B_{x, y}}
\end{equation}
where $B_{x, y}$ is the brightness at pixel located at $\left(x, y\right)$, and $A_{intersection}$ and $A_{union}$ are, respectively, the set of pixels inside the intersection and union of the predicted/ground-truth blob mask.
\end{itemize}

\phantomsection
\section*{Results}
\label{sec:results}
We present results in four parts: (1) training results for each model described in \hyperref[subsec:training]{Training}, (2) testing scores of the trained models on both synthetic and real GPI data shown in \hyperref[subsec:testing]{Testing and blob tracking}, (3) furthermore, the models are evaluated based on human-labeled blobs to demonstrate their validity in \hyperref[subsec:humanlabel]{Evaluation by human-labeled blobs}, (4) the tracking information from each model is used to identify the regime of the blob dynamics in \hyperref[subsec:regime]{Identification of blob regimes} to perform a task that addresses an active plasma-physics research topic.

\phantomsection
\subsection*{Training}
\label{subsec:training}

We train the tracking models using 30 synthetic videos, each with 200 frames. We split the frames into training data (95 percent) and validation data (5 percent). We used a training pipeline similar to the original implementation with default hyperparameters for RAFT and GMA. For Mask R-CNN, we find hyperparameters by Bayesian optimization (see the supplementary information). For Flow Walk, we implemented the training pipeline from RAFT. The training scores based on the \hyperref[subsubsec:metrics]{Performance metrics} are shown in the supplementary information, where RAFT is the best among the four models.

\phantomsection
\subsection*{Testing and blob tracking}
\label{subsec:testing}
We test the models on synthetic and real GPI data and summarize the results in Table \ref{tab:result_testing}. For the real GPI data, we only compute VIoU since there are no labels in optical flows. Among the four models, RAFT performs best on real data.

\begin{table}[ht]
\phantomsection
\centering
\caption{Scores from testing for each model with corresponding score metrics, endpoint error (EPE) and volumetric IoU (VIoU). Lower EPE and higher VIoU are better. Among four models, RAFT performs best with the lowest EPE and highest VIoU. Mask R-CNN misses EPE because it is not an optical flow detection model but a mask detection model. EPE for real data is not presented because the real data does not have ground-truth optical flow velocities.}
\begin{tabular}{|l|ll|ll|}
\hline
                             & \multicolumn{2}{c|}{\begin{tabular}[c]{@{}c@{}}Testing score\\ on synthetic data\end{tabular}} & \multicolumn{2}{c|}{\begin{tabular}[c]{@{}c@{}}Testing score\\ on real data\end{tabular}} \\ \hline
Models\textbackslash{}Metric & \multicolumn{1}{l|}{EPE}                                 & VIoU                                & \multicolumn{2}{l|}{VIoU}                                                                 \\ \hline
RAFT                         & \multicolumn{1}{l|}{2.701}                               & 0.783                               & \multicolumn{2}{l|}{0.781}                                                                \\ \hline
Mask R-CNN                   & \multicolumn{1}{l|}{N/A}                                   & 0.638                               & \multicolumn{2}{l|}{0.700}                                                                \\ \hline
GMA                          & \multicolumn{1}{l|}{2.974}                               & 0.777                               & \multicolumn{2}{l|}{0.650}                                                                \\ \hline
Flow Walk                    & \multicolumn{1}{l|}{6.020}                               & 0.631                               & \multicolumn{2}{l|}{0.621}                                                                \\ \hline
\end{tabular}
\label{tab:result_testing}
\end{table}

\begin{figure}[ht]
  \phantomsection
  \centering
  \includegraphics[width=1.0\textwidth]{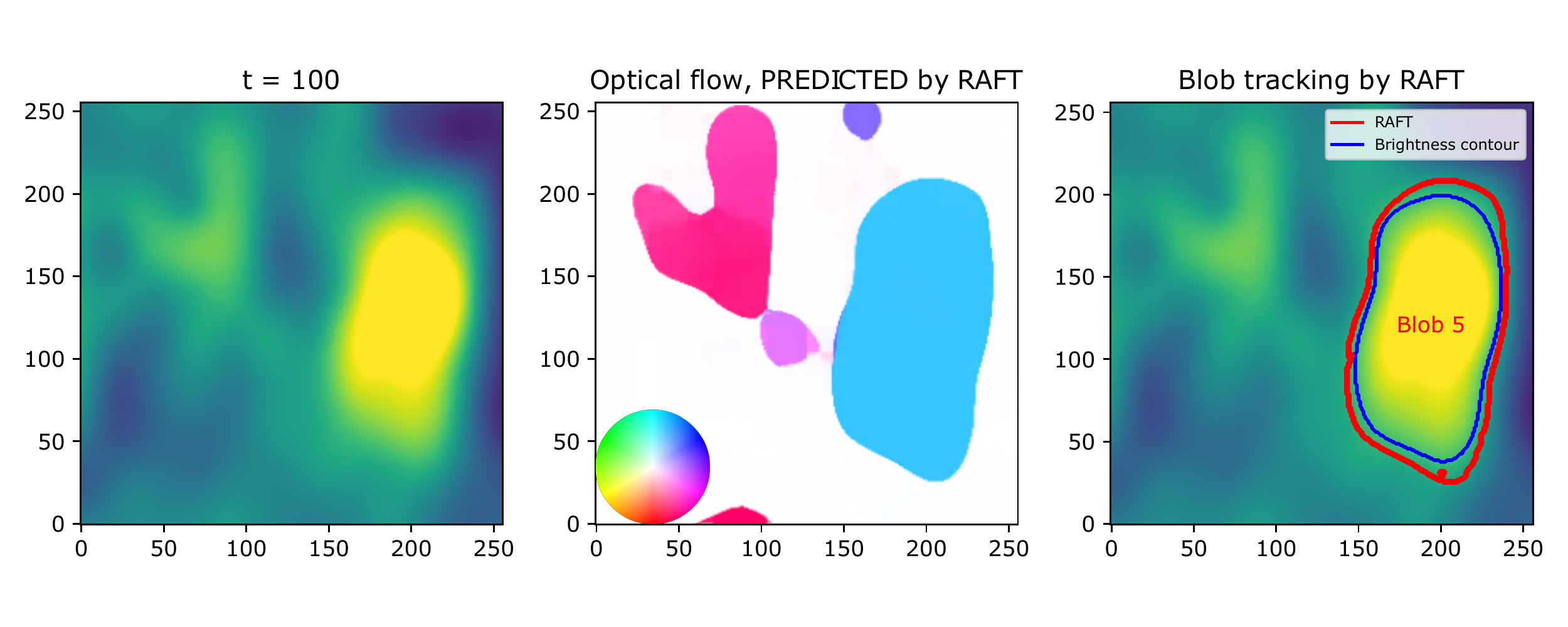}
  \caption{An example of blob tracking on experimental data. With the current frame (left) and the next frame as input, RAFT predicts optical flows for each pixel (middle). In the middle, the color of each pixel indicates the velocity (speed and direction) of the optical flow predicted, corresponding to the color palette shown at the bottom left. On the right, the red contour is drawn for the boundary of the pixels having non-zero optical flow in the central figure. The blue contour is the contour of $0.7\times B_{max}$ where $B_{max}$ is the maximum (standardized) brightness inside the red contour, which is used for computing VIoU. The red contours chosen by several thresholds (VIoU $> 0.8$, $B_{max}>0.75$, lifetime $> 15$ frames) are only tracked. The blob ID (Blob 5) was assigned by \hyperref[subsubsec:tbdalgo]{Tracking-by-detection}.}
  \label{fig:raft_tracking_example}
\end{figure}

Figure \ref{fig:raft_tracking_example} shows an example of blob tracking by RAFT on real experimental data. RAFT takes two images as input: the current and the next frame. It then predicts optical flows for each pixel, as shown in the middle of Figure \ref{fig:raft_tracking_example}. The color of each pixel indicates the velocity of the predicted optical flow, corresponding to the color palette shown at the bottom left. In order to identify the mask of the blobs from the map of optical flows, which is not segmented, the algorithm computes a set of contours of different values of the magnitude of optical flows in each frame. We merge contours that overlap within VIoU $>\alpha$. We find $\alpha = 0.2$ suitable for our purposes. For each contour, we compute the VIoU using the blue contour in the figure, which is the contour of $\beta \times B_{max}$, where $B_{max}$ is the maximum (standardized) brightness inside the red contour. Then the contours which give a VIoU greater than $\gamma = 0.8$ and $B_{max}$ greater than $\beta = 0.7$ are chosen to be tracked. The user sets the hyperparameters $\alpha$, $\beta$, and $\gamma$ based on the GPI data, and we set them as above for the data used in this work. In Figure \ref{fig:raft_tracking_example} (right), the blob ID (Blob 5) is assigned by the \hyperref[subsubsec:tbdalgo]{Tracking-by-detection}. In order to enforce temporal coherence in blob detection, only blobs with a lifetime longer than 15 frames (corresponding to 7.5 micro-seconds in real data) are chosen and drawn in the output video. Unlike the optical flow prediction models, Mask R-CNN outputs segmented masks of objects, and therefore the contour of an object is already obtained (i.e., we do not need to scan the values of contours). The rest of the tracking is the same as above.

\begin{figure}[ht]
  \phantomsection
  \centering
  \includegraphics[width=1.0\textwidth]{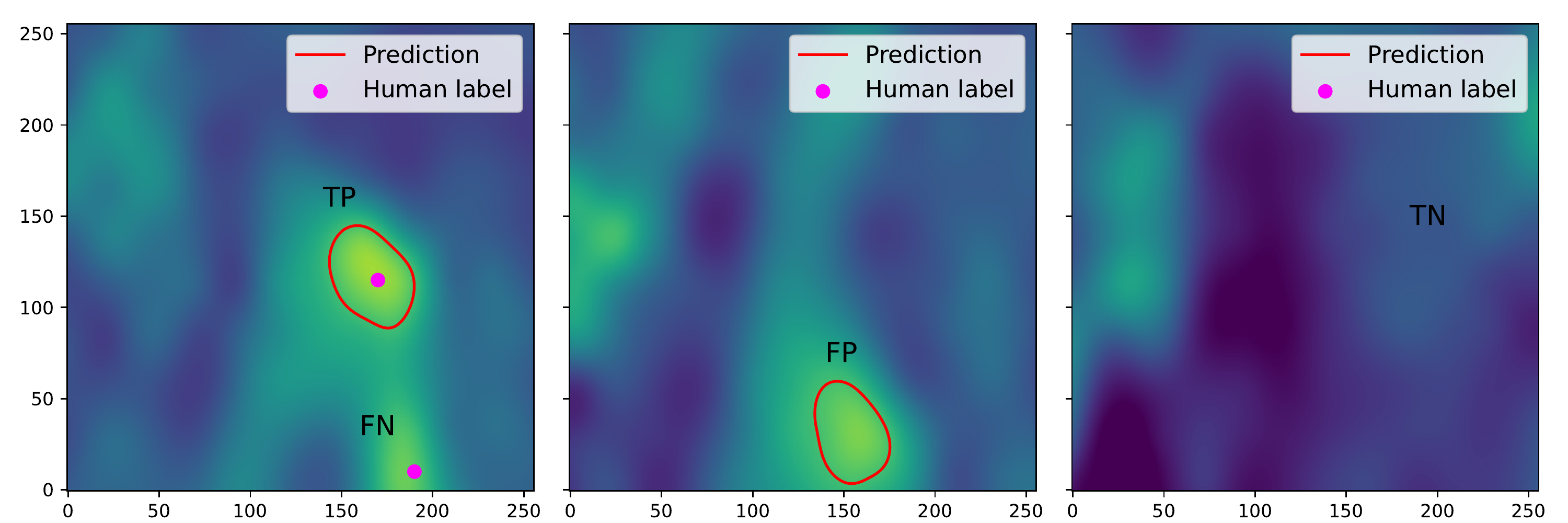}
  \caption{Examples of model prediction on the real data with human labels. The labeler marks a dot (magenta) on the blob that he/she identifies in every frame. As illustrated in the left figure, TP is true positive where the human-labeled dot is contained inside the predicted blob boundary, and FN is false negative where the human-labeled dot is not contained in any predicted blob boundary. FP is a false positive (middle figure) where the predicted blob boundary does not contain human labels. In the right figure, there are neither predictions nor human labels in this frame, hence the case of true negative.}
  \label{fig:hand_label_example}
\end{figure}

\phantomsection
\subsection*{Evaluation by human-labeled blobs}
\label{subsec:humanlabel}

\begin{figure}[ht]
  \phantomsection
  \centering
  \includegraphics[width=1.0\textwidth]{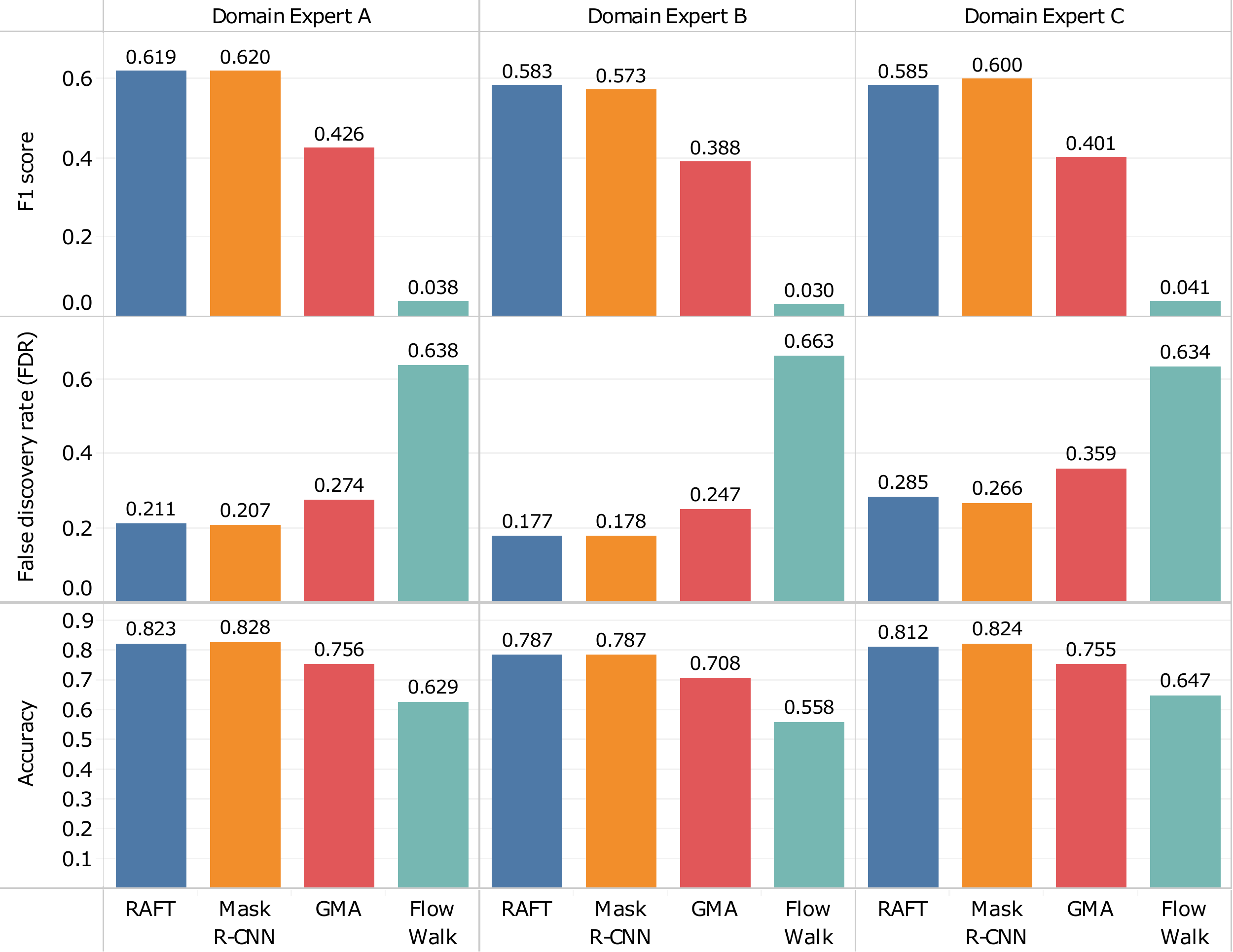}
  \caption{F1 score, false discovery rate (FDR), and accuracy of the four methods (RAFT, Mask R-CNN, GMA, and Flow Walk) on the real GPI data with 3,000 frames hand-labeled by three domain experts (A, B, and C). Metrics shown are the average per-frame scores (i.e., the average across the frames).}
  \label{fig:hand_labels}
\end{figure}

\begin{figure}[ht]
  \phantomsection
  \centering
  \includegraphics[width=1.0\textwidth]{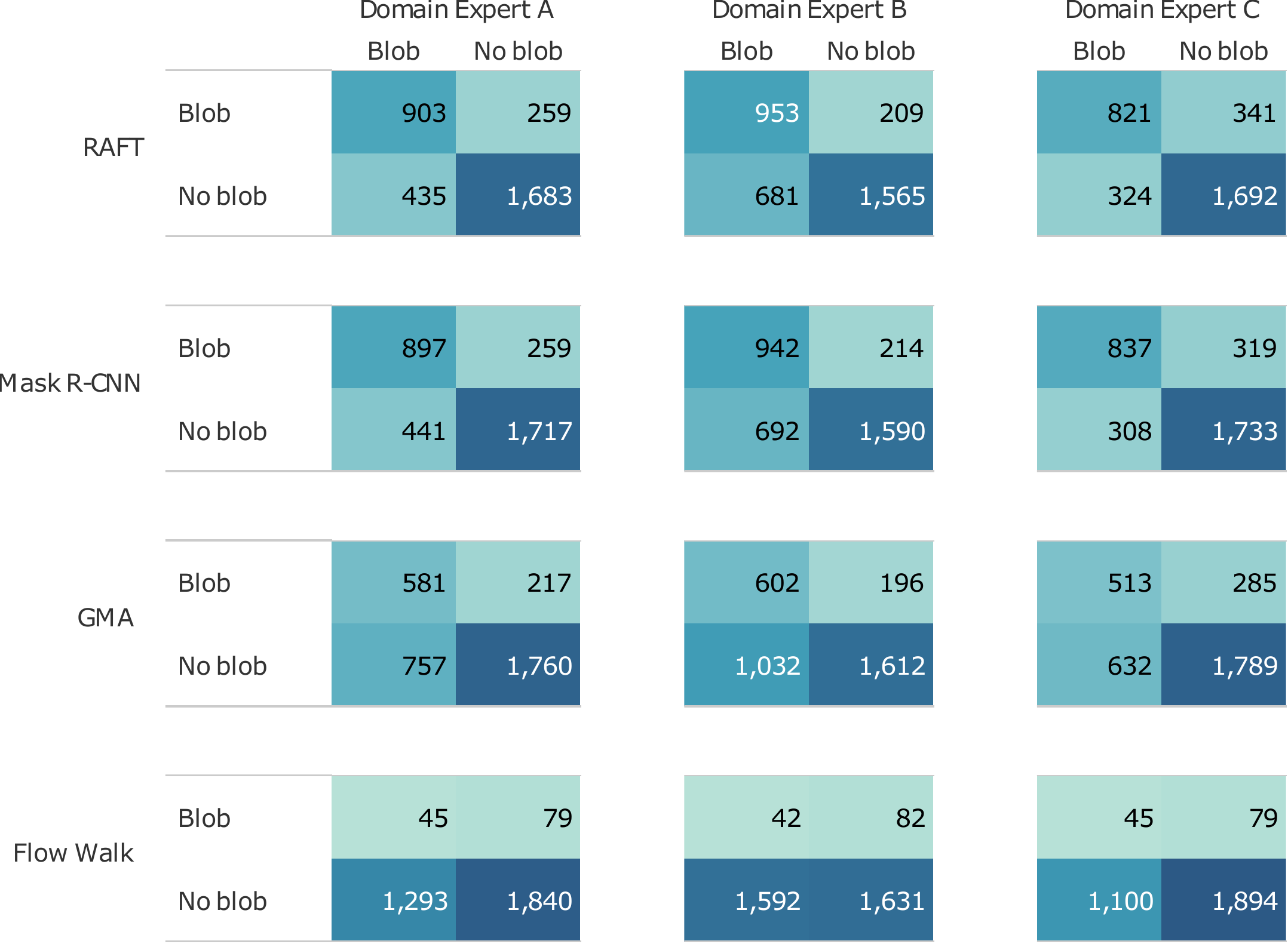}
  \caption{Confusion matrices of the four methods (RAFT, Mask R-CNN, GMA, and Flow Walk), for the real GPI data with 3,000 frames hand-labeled by three domain experts (A, B, and C).}
  \label{fig:conf_matrix}
\end{figure}

For the testing score on real data shown in Table \ref{tab:result_testing}, the label (blue contour in Figure \ref{fig:raft_tracking_example} (right)) is dependent on the prediction (red contour). In other words, the label is drawn only for the predicted objects, and the model performance is thus evaluated only for structures detected by our tracking models (i.e., there are only true positives). In general, there is some subjectivity in identifying blobs in real data. Therefore, we quantified how close the machine-predicted blobs are to the human-labeled blobs selected by different domain experts. We consider as ``ground-truth" the cases that human labelers have identified as blobs. We cover cases with false positives (the model identified a blob where the human identified none), true negatives (did not identify a blob where there was none), false negatives (did not identify a blob where there was one), as well as true positives (identified a blob where there was one), as defined in Figure \ref{fig:hand_label_example}. Each of the three domain experts separately labeled the blobs in 3,000 frames by hand, and our blob-tracking models are evaluated against these human-labeled experimental data based on F1 score, False Discovery Rate (FDR), and accuracy, as shown in Figure \ref{fig:hand_labels}. These are the average per-frame scores (i.e., the average across the frames), and we did not use the score across all frames, which can be dominated by outlier frames that may contain many blobs. Figure \ref{fig:conf_matrix} displays the corresponding confusion matrices. In this result, RAFT, Mask R-CNN, and GMA achieved high accuracy (0.807, 0.813, and 0.740 on average, respectively), while Flow Walk was less accurate (0.611 on average). Here, the accuracy of 0.611 in Flow Walk is seemingly high, misleading because Flow Walk gave few predictions (low TP and FP in Figure \ref{fig:conf_matrix}). This is because the data is skewed to true negatives as many frames have no blobs, which is seen from the high true negatives of confusion matrices in Figure \ref{fig:conf_matrix}. Thus, accuracy is not the best metric for the data used. F1 score and FDR are more suitable for our purposes because they are independent of true negatives. Indeed, other scores of Flow Walk are as expected; the F1 score is low (0.036 on average) and the FDR is high (0.645 on average). RAFT and Mask R-CNN show decently high F1 scores and low FDR. GMA underperformed RAFT and Mask R-CNN in all metrics, but the scores are fairly good. These observations are consistent across the comparisons with the three labelers' results. Overall, blobs predicted by three models (RAFT, Mask R-CNN, and GMA) are quite similar to human-labeled blobs, exhibiting fairly good scores in Figure \ref{fig:hand_labels}.

The trained models' average running times per frame (without post-processing) are milliseconds. Specifically, 104, 50, 109, and 23 milliseconds for RAFT, Mask R-CNN, GMA, and Flow Walk, respectively, using 4 GPUs. For RAFT and GMA, there are 12 iterations for each frame following the default setup, and we run 1 iteration for Mask R-CNN and Flow Walk. This indicates that $\sim3$ hours (RAFT, GMA) and $\sim1$ hours (Mask R-CNN, Flow Walk) would be needed to process $10^5$ images. This can become faster if we reduce the number of iterations or downsample the input image at the expense of the model's performance. The post-processing takes seconds, precisely 12, 11, and 9 seconds per frame for RAFT, GMA, and Flow Walk, respectively, and $\sim300$ hours for $10^5$ images. This is not short because of the search for the contour value to be drawn from the optical flow for the blob contour prediction. For Mask R-CNN, the per-frame post-processing time is 1 second, much faster than the other three optical flow models. This is because Mask R-CNN gives segmented masks, and there is no need to find a contour value. Therefore, RAFT, GMA, and Flow Walk can be run for a small subset of the data ($\sim$1,000 frames), and Mask R-CNN can be run for the data with a larger number of frames. Despite the long computation time, the post-processing provides information of blobs for every frame, whereas CAS only provides average results in a relatively shorter time.

\phantomsection
\subsection*{Identification of blob regimes}
\label{subsec:regime}
\begin{figure}[ht]
  \phantomsection
  \centering
  \includegraphics[width=1.0\textwidth]{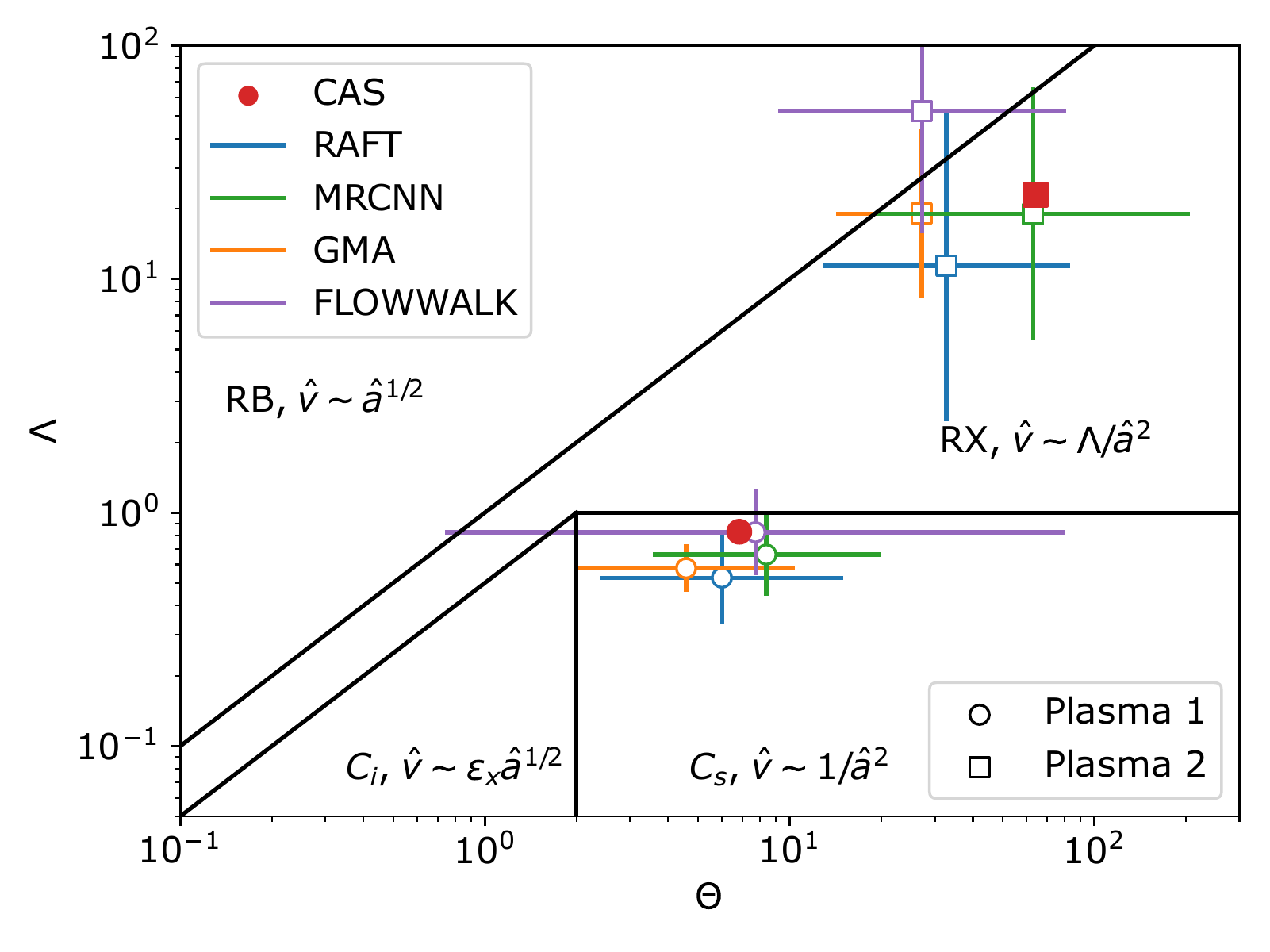}
  \caption{$\Theta$---$\Lambda$ diagram where $\Theta$ and $\Lambda$ are defined in Equation \ref{eq:theta_lambda}. The diagram is divided into four blob regimes (RB, RX, $C_i$, and $C_s$) depending on the relation between the normalized speed ($\hat{v}$) and radius ($\hat{a}$) of the blobs. Data from two different plasmas (Plasma 1 and Plasma 2) are interrogated by four blob-tracking models (RAFT, Mask R-CNN, GMA, and Flow Walk) and the traditional CAS method, with the locations from each method indicated in the diagram. For these data, the factor $\epsilon_{x}$ is 0.5. The bars correspond to the standard deviation of $\log{\Theta}$ and $\log{\Lambda}$.}
  \label{fig:myra_result}
\end{figure}

The trained models can now estimate various blob parameters, such as their size, speed, and occurrence frequency in real GPI data. Here, we use such information to identify the regime of the blob dynamics for two different plasma conditions. The regime is identified based on location in the diagram in Figure \ref{fig:myra_result}. Here, $\Theta$ (a normalized blob size) and $\Lambda$ (a normalized plasma collisionality, which is proportional to the electron-ion collision frequency) are defined as in \cite{Myra_2006_1,Myra_2006_2}
\begin{equation}
\phantomsection
\label{eq:theta_lambda}
\Theta=\hat{a}^{5/2}
\quad\mathrm{and}\quad
\Lambda=1.7\times10^{-18}\frac{n_{e}L_{\|}}{T_{e}^2}
\end{equation}
where
\begin{equation}
\phantomsection
\label{eq:a_v}
\hat{a}=\frac{a_{b}R^{1/5}}{L_{\|}^{2/5}\rho_{s}^{4/5}}
\quad\mathrm{and}\quad
\hat{v}=\frac{v_{R}}{c_{s}\left(2L_{\|}\frac{\rho_{s}^2}{R^3}\right)^{1/5}}
\end{equation}
$\hat{a}$ and $\hat{v}$ are the blob's normalized radius and radial speed, respectively. Here, $n_e$ and $T_e$ are the local electron density and temperature, respectively, $L_{\|}$ is the parallel connection length, $a_b$ is the blob radius, $R$ is the major radius of the tokamak, $\rho_{s}$ is the ion sound Larmor radius, $v_R$ is the radial speed of the blob, and $c_s$ is the sound speed. Our blob-tracking models allow us to estimate $a_b$ and $v_R$. Other plasma parameters are measured using other diagnostics. There are four regimes in the diagram, named ``resistive ballooning" (RB), ``resistive X-point" (RX), ``connected ideal-interchange" ($C_i$), and ``connected sheath" ($C_s$). Theory predicts different scaling relationships between the normalized radial speed ($\hat{v}$) and the radius ($\hat{a}$) of the blob depending upon  regime (see Figure \ref{fig:myra_result}). We have evaluated $\Theta$
and $\Lambda$ for two different plasma conditions (Plasma 1 and Plasma 2) using a traditional method (CAS) and our four blob-tracking models (RAFT, Mask R-CNN, GMA, and Flow Walk). This evaluation locates the two plasmas within the theory-defined regimes, as shown in Figure \ref{fig:myra_result}. We identify plasma 1 and 2 as being within $C_s$ and RX regimes, respectively, unanimously by four methods (CAS, RAFT, Mask R-CNN, and GMA) except Flow Walk. The closeness of the data points between the traditional method and the blob-tracking models demonstrates the validity of the machine learning approach in blob-tracking applied to an important research investigation. Note that the bars around the centroids for the blob-tracking models are not error bars but rather the spread in actual blob statistics. This is an advantage of the blob-tracking methods since they allow statistics derived from individual blob measurements, whereas CAS can only yield average results. Flow Walk's large spread is due to its poor prediction performance, as shown previously in Figure \ref{fig:hand_labels}.

\phantomsection
\section*{Conclusions}
\label{sec:conclusion}
We introduce a dataset and benchmark for tracking turbulent structures, called blobs, in Gas Puff Imaging data from the boundary region of magnetically confined tokamak plasmas created in TCV.
We experimented with four baseline models based on RAFT, Mask R-CNN, GMA, and Flow Walk, trained on synthetic data, and tested on synthetic and real data. The synthetic data mimic the realistic movement and evolution of blobs found in actual GPI videos of the TCV boundary plasma. We optimize the training hyperparameters by Bayesian optimization. The trained models show high VIoU testing scores on both synthetic and real data and high accuracy scores on human-labeled real data. As a demonstration of the validity of our tracking models, we achieve (nearly) unanimous identification of the regime of the blob dynamics for two different plasma conditions using blob statistics estimated by tracking models and the traditional CAS method. At least two of these models (RAFT and Mask R-CNN) are reliable tools for identifying and tracking blobs, which allow the estimation of various measures of blob dynamics connected to the levels and effects of turbulence on the edge of tokamak plasmas.
This characterization of edge turbulence furthers our understanding of transport processes in the plasma boundary, contributing to essential knowledge for the practical generation of fusion energy.

\section*{Data availability}
In the spirit of reproducible research, we make our data, models, and code publicly available at \url{https://github.com/harryh5427/GPI-blob-tracking}.

\section*{Acknowledgements}
The support from the US Department of Energy, Fusion Energy Sciences, awards DE-SC0014264 and DE-SC0020327, are gratefully acknowledged. This work was supported in part by the Swiss National Science Foundation. Also, this work has been carried out within the framework of the EUROfusion Consortium, funded by the European Union via the Euratom Research and Training Programme (Grant Agreement No 101052200 -EUROfusion). Views and opinions expressed are, however those of the author(s) only and do not necessarily reflect those of the European Union or the European Commission. Neither the European Union nor the European Commission can be held responsible for them.

\section*{Author contributions}
W.H. constructed the initial concept of the study, curated the dataset, and wrote the code. W.H. and I.D. defined the benchmark and reviewed the code. All authors provided ideas and analyzed the results. W.H., R.A.P., R.V.L., N.O., T.G., C.T., J.T., and I.D. contributed to the writing and all authors reviewed the manuscript. W.H., I.D., and R.V.L. created figures.

\bibliography{bibliography}

\newpage
\clearpage

\section*{Appendix}
\begin{figure}[ht]
  \centering
  \includegraphics[width=1.0\textwidth]{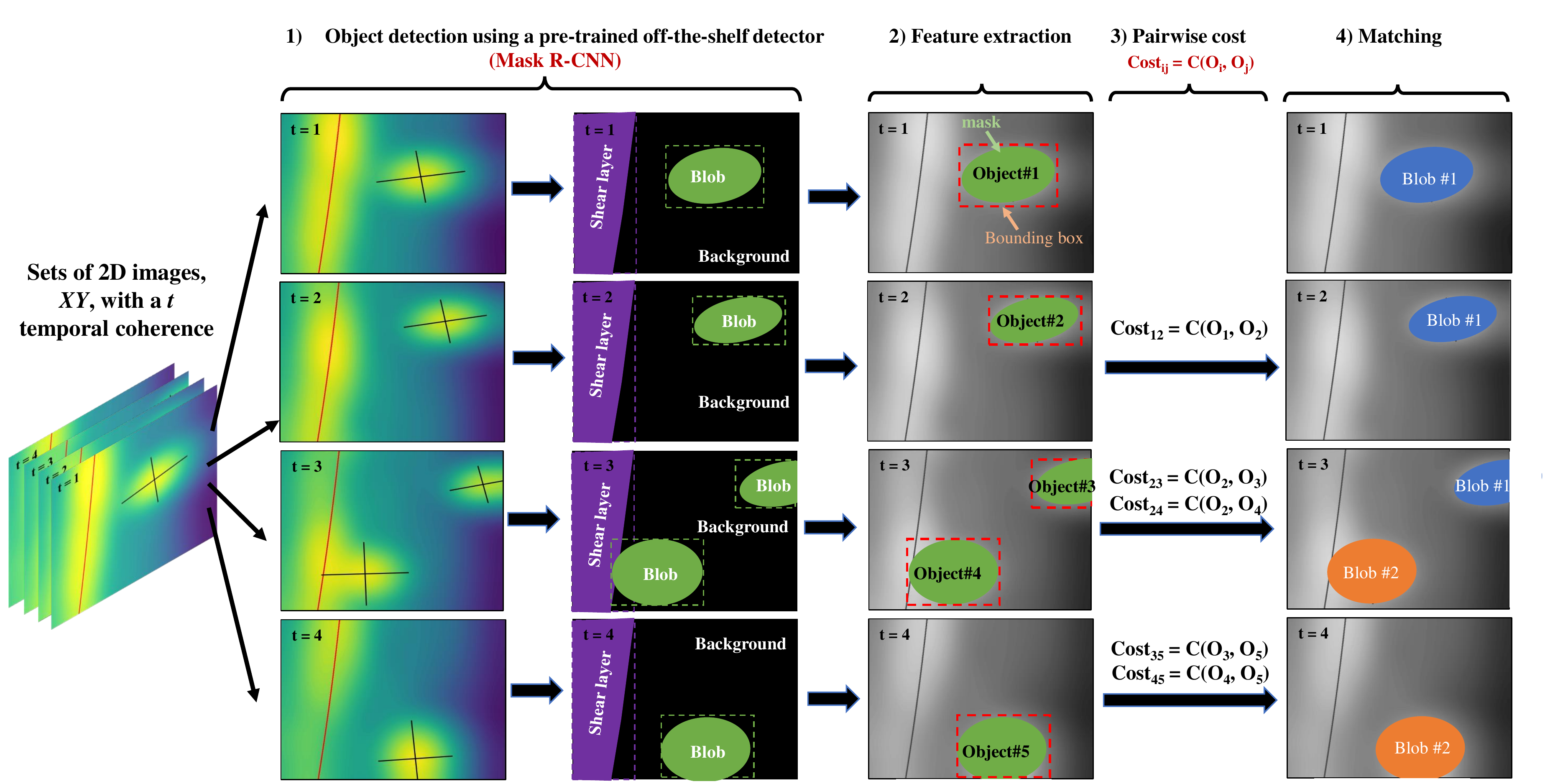}
  \caption{\hyperref[subsubsec:tbdalgo]{Tracking-by-detection} consisting of four steps: (1) Object detection within each frame using a pre-trained model. Here the mask R-CNN produces a bounding box and a mask for each candidate object. (2) Extraction of features of interest (i.e., masks of blobs). (3) Computation of pairwise costs between objects in the current and previous frame. (4) Bipartite matching between objects assigning unique correspondence.}
  \label{fig:tracking_schematic}
\end{figure}

\begin{figure}[ht]
  \centering
  \includegraphics[width=1.0\textwidth]{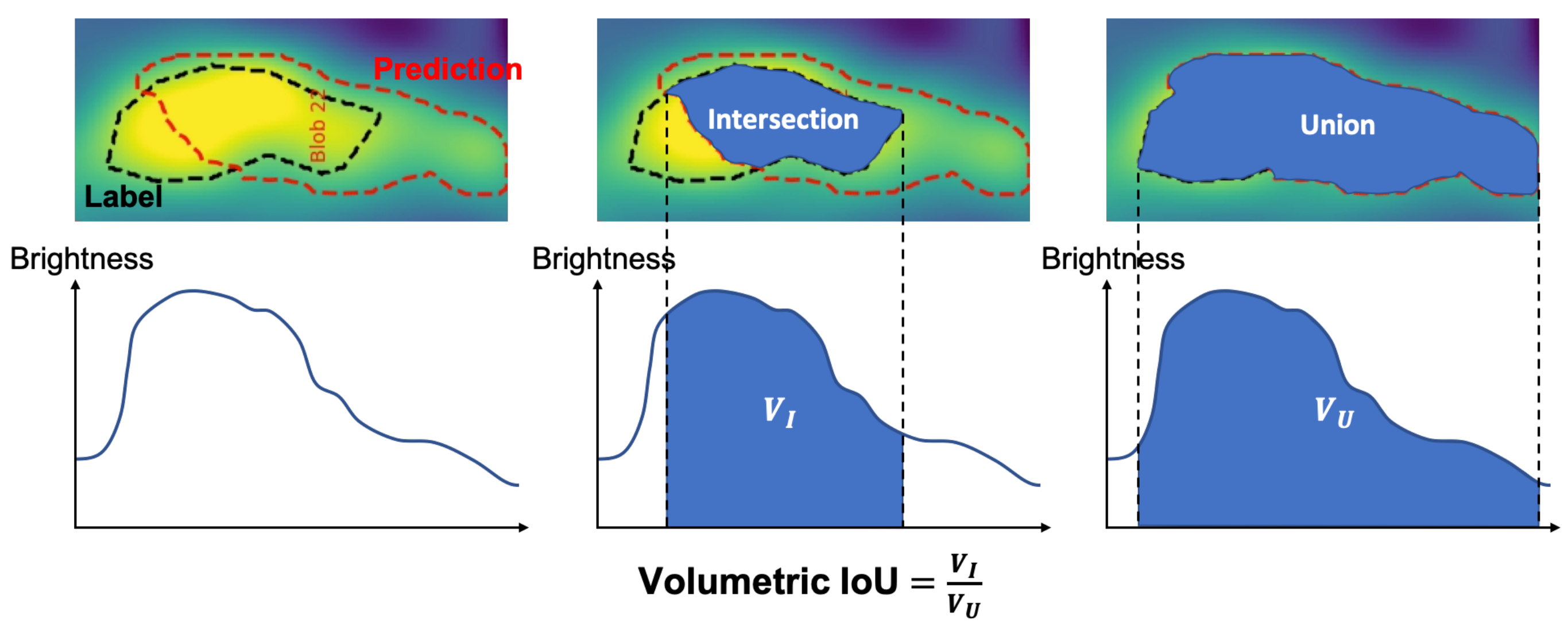}
  \caption{Volumetric Intersection over Union (VIoU) computed for a prediction on an image containing a blob (top) using the volume under the profile of the brightness (bottom).}
  \label{fig:volumetric_iou}
\end{figure}

We provide additional details on the methods and results. Figure \ref{fig:tracking_schematic} illustrates \hyperref[subsubsec:tbdalgo]{Tracking-by-detection} workflow, which is applied to the consecutive frames containing blob contours predicted by the model and assigns blob IDs based on closeness to the blobs in the previous frame. Figure \ref{fig:volumetric_iou} shows an example of computing the VIoU between the contours of a prediction (red) and a label (black). For the training of Mask R-CNN, we explore the hyperparameters in Table \ref{tab:hyperparameters_range} by Bayesian optimization and find the optimal values shown in Table \ref{tab:BO_result}. The scores of the four models on training and validation synthetic data are shown in Table \ref{tab:result_training} with corresponding score metrics (EPE and VIoU). The EPE scores of Mask R-CNN are not applicable because it is not an optical flow detection model but a mask detection model.

\begin{table}[ht]
\begin{center}
\begin{minipage}{\textwidth}
\caption{Exploration ranges of hyperparameters used in Bayesian optimization. BO level 1 hyperparameters: learning rate initial value, learning rate reduction period, learning rate decay factor, momentum, weight decay and number of training epochs. BO level 2 hyperparameters: probabilities for horizontal flip, scaling, translation, shearing, rotation, and dropout.}
\label{tab:hyperparameters_range}
\begin{tabular*}{\textwidth}{@{}l|c@{\hspace{0.33cm}}c@{\hspace{0.33cm}}c@{\hspace{0.33cm}}c@{\hspace{0.33cm}}c@{\hspace{0.33cm}}c@{}}
\toprule%
BO Lv. 1 & \vtop{\hbox{\strut LR}\hbox{\strut initial val}} & \vtop{\hbox{\strut LR}\hbox{\strut reduction}\hbox{\strut period}} & \vtop{\hbox{\strut LR}\hbox{\strut decay}} & Momentum & \vtop{\hbox{\strut Weight}\hbox{\strut decay}} & $n_{epochs}$ \\
\hline
\begin{tabular}[c]{@{}l@{}}Exploration\\ range\end{tabular} & 1e-6---5e-2 & 1---40 & 0.1---1.0 & 0.0---1.0 & 0.0---1e-3 & 2---40 \\
\midrule
BO Lv. 2 & $P_{horizontalFlip}$ & $P_{scale}$ & $P_{translate}$ & $P_{shear}$ & $P_{rotate}$ & $P_{dropout}$ \\
\hline
\begin{tabular}[c]{@{}l@{}}Exploration\\ range\end{tabular} & 0.0---1.0 & 0.0---1.0 & 0.0---1.0 & 0.0---1.0 & 0.0---1.0 & 0.0---0.5 \\
\midrule
\end{tabular*}
\end{minipage}
\end{center}
\end{table}

\begin{table}[ht]
\begin{center}
\begin{minipage}{\textwidth}
\caption{Optimal values of hyperparameters found from Bayesian optimization.}
\label{tab:BO_result}
\begin{tabular*}{\textwidth}{@{}l|c@{\hspace{0.33cm}}c@{\hspace{0.33cm}}c@{\hspace{0.33cm}}c@{\hspace{0.33cm}}c@{\hspace{0.33cm}}c@{}}
\toprule%
BO Lv. 1 & \vtop{\hbox{\strut LR}\hbox{\strut initial val}} & \vtop{\hbox{\strut LR}\hbox{\strut reduction}\hbox{\strut period}} & \vtop{\hbox{\strut LR}\hbox{\strut decay}} & Momentum & \vtop{\hbox{\strut Weight}\hbox{\strut decay}} & $n_{epochs}$ \\
\hline
\begin{tabular}[c]{@{}l@{}}Optimal\\ value\end{tabular} & 0.050 & 12 & 0.402 & 0.296 & 2.729e-5 & 28 \\
\midrule
BO Lv. 2 & $P_{horizontalFlip}$ & $P_{scale}$ & $P_{translate}$ & $P_{shear}$ & $P_{rotate}$ & $P_{dropout}$ \\
\hline
\begin{tabular}[c]{@{}l@{}}Optimal\\ value\end{tabular} & 0.341 & 0.560 & 0.261 & 0.564 & 0.368 & 1.283e-17 \\
\midrule
\end{tabular*}
\end{minipage}
\end{center}
\end{table}

\begin{table}[ht]
\centering
\caption{Scores from training and validation for each model with corresponding score metrics, endpoint error (EPE) and volumetric IoU (VIoU). Mask R-CNN misses EPE because it is not an optical flow detection model but a mask detection model.}
\begin{tabular}{|l|ll|ll|}
\hline
                             & \multicolumn{2}{c|}{\begin{tabular}[c]{@{}c@{}}Training score\\ on synthetic data\end{tabular}} & \multicolumn{2}{c|}{\begin{tabular}[c]{@{}c@{}}Validation score\\ on synthetic data\end{tabular}} \\ \hline
Models\textbackslash{}Metric & \multicolumn{1}{l|}{EPE}    & VIoU  & \multicolumn{1}{l|}{EPE}     & VIoU   \\ \hline
RAFT                         & \multicolumn{1}{l|}{0.161}  & 0.869 & \multicolumn{1}{l|}{0.338}   & 0.879  \\ \hline
GMA                          & \multicolumn{1}{l|}{0.160}  & 0.865 & \multicolumn{1}{l|}{0.298}   & 0.862  \\ \hline
Mask R-CNN                   & \multicolumn{1}{l|}{N/A}      & 0.705 & \multicolumn{1}{l|}{N/A}       & 0.704  \\ \hline
Flow Walk                    & \multicolumn{1}{l|}{0.946}  & 0.835 & \multicolumn{1}{l|}{0.924}   & 0.830  \\ \hline
\end{tabular}
\label{tab:result_training}
\end{table}

\end{document}